# Pressure-induced superconductivity in charge-density-wave compound LaTe$_{2-x}$Sb$_x$ (x=0 and 0.4)

Xu Chen,[1,†] Pei-han Sun,[2,†] Zhenkai Xie,[1,3] Fanqi Meng,[1] Cuiying Pei,[4] Yanpeng Qi,[4,5,6] Tianping Ying,[1] Kai Liu,[2,*] Jian-gang Guo,[1,7,*] and Xiaolong Chen[1,3,*]

1. *Beijing National Laboratory for Condensed Matter Physics, Institute of Physics, Chinese Academy of Sciences, Beijing 100190, China*
2. *Department of Physics and Beijing Key Laboratory of Opto-electronic Functional Materials & Micro-nano Devices, Renmin University of China, Beijing 100872, China*
3. *School of Physical Sciences, University of Chinese Academy of Sciences, Beijing 100049, China*
4. *School of Physical Science and Technology, ShanghaiTech University, Shanghai 201210, China*
5. *ShanghaiTech Laboratory for Topological Physics, ShanghaiTech University, Shanghai 201210, China*
6. *Shanghai Key Laboratory of High-resolution Electron Microscopy and ShanghaiTech Laboratory for Topological Physics, ShanghaiTech University, Shanghai 201210, China*
7. *Songshan Lake Materials Laboratory, Dongguan, Guangdong 523808, China*

[†]These authors contributed equally
[*]kliu@ruc.edu.cn, jgguo@iphy.ac.cn, chenx29@iphy.ac.cn

## Abstract

Here, we have grown single crystals of LaTe$_{2-x}$Sb$_x$ (x=0 and 0.4) with continuously adjustable CDW. High-pressure x-ray diffraction show LaTe$_2$ does not undergo phase transition and keep robust below 40 GPa. *In-situ* high-pressure electrical measurements show LaTe$_{2-x}$Sb$_x$ undergo semiconductor-metal-superconductivity transition at 4.6 and 2.5 GPa, respectively. With the doping of Sb, the highest $T_c$ increases from 4.6 to 6.5 K. Theoretical calculations reveal that the CDW has been completely suppressed and the calculated $T_c$ is about 2.97 K at 4.5 GPa, consistent with the measured value. Then, the pressure-induced superconductivity in LaTe$_{2-x}$Sb$_x$ can be explained in the framework of the BCS theory.

## Introduction

Layered compounds have been research hotspots due to their sensitively pressure-dependent structure and electronic property.[1,2,3] As the interlayer spacing shrinks, the structure tends to be three dimensionality and electron becomes more conductivity because of enhanced overlap of atomic orbitals.[4,5,6,7] In some layered materials, application of chemical or physical pressure tunes the ground states from commensurate CDW to normal metal and SC.[8,9] However, chemical pressure inevitably introduces defects or atomic disorders, undermining the effect of lattice contraction on the physical property. Physical pressure avoids the complexity brought by chemical doping, providing a clean and powerful route for continuous tuning lattice structure and the corresponding electronic states. It has been widely utilized in layered materials to discern the interaction between CDW and SC.[10,11,12,13]

In transition metal dichalcogenides (TMDCs), such as 1$T$-TaS$_2$ and 1$T$-TaSe$_2$ with a Mott-insulating state and multiple CDW phases, the superconducting transition temperature ($T_c$) under pressure remains increases regardless of the change in the CDW, indicating CDW is weakly connected to SC.[9,14,15] In other TMDCs like 1$T$-TiSe$_2$, 2$H$-TaS$_2$ and 2$H$-TaSe$_2$, the dome-shaped superconducting phase diagrams suggest that the SC directly competes with CDW.[16,17,18] Obviously, *in-situ* pressure broadens our views of the structural and electronic properties, and more intriguing phenomena in layered materials are highly expected through pressure regulation.

RETe$_2$, contain CDW at higher temperatures, show layered tetragonal Cu$_2$Sb-type structure, where Te(1) atoms form the Te square sheet that is sandwiched by the corrugated double layers of RE-Te(2) slabs along the *c* axis.[19,20] Remarkably, CeTe$_2$ is the first pressure-induced superconductor discovered in RETe$_2$ family. It undergoes two different magnetic orderings (the short-range ferromagnetic ordering temperature $T_{SRF}$ and the long-range antiferromagnetic ordering temperature $T_N$) at relatively low temperatures.[21] With increasing pressure, the SC suddenly appears below 2 kbar and slightly increases, along with enhances both $T_{SRF}$ and $T_N$ over the whole region of measured pressure. Although the superconductivity seems likely to be related to magnetic fluctuation, the details were still unclear.[22] Subsequent theoretical work suggest that both Te vacancy and pressure can induce the charge transfer and increased Te(1) 5$p$ hole carriers are responsible for the pressure-induced superconductivity in CeTe$_2$. However, the complex magnetic orders make the role of CDW and superconductivity mechanism still difficult to understand.[23]

The isostructural compound, LaTe$_2$, avoids magnetic order of rare earths, is more suitable for research on the relationship between SC and CDW in RETe$_2$ family than CeTe$_2$.[24,25] It was reported to have the CDW instability along [100] direction in the *ab* plane, raised from the Fermi-surface nesting between the electron and hole Fermi surfaces in the Te(1) square sheet.[26] The theoretical work expect that the additional pressure effect would enhance the superconductivity in LaTe$_2$. On the other hand, the correspondence between the geometry of Fermi surfaces and the superlattice wave vectors is observed in the LaTe$_{2-x}$Sb$_x$ system, this distortions are due to charge-density wave driven by the considerable nesting of the very simple square-sheet band structure.[27] Accordingly, non-magnetic LaTe$_{2-x}$Sb$_x$ with continuously adjustable CDW could provide a good platform for research on the origin of pressure-induced superconductivity in RETe$_2$ family.

Herein, we have studied the high-pressure transport properties in LaTe$_{2-x}$Sb$_x$ ($x$=0 and 0.4). With increasing pressure, LaTe$_{2-x}$Sb$_x$ undergoes a semiconductor-metal-superconductivity transition, and the maximal $T_c$ are 4.6 K and 6.5 K, respectively. Theoretical calculations suggest that the CDW has been completely suppressed at 4.5 GPa, and electron-phonon coupling is responsible for pressure-induced SC in RETe$_2$ systems, rather than magnetism.

## Methods

**Sample preparation**

Black platelike-shaped single crystals of LaTe$_{2-x}$Sb$_x$ were obtained by chemical vapor transport reactions in quartz tube with I$_2$ as the transporting agent. The starting materials are La ingot (Alfa, 99.99%), Te pieces (Alfa, 99.999%) and Sb grains (Alfa, 99.999%), as the recipe reported earlier.[19] The high and low temperatures are set as 1220 K and 1120 K, respectively. The obtained crystals were kept in a dry and oxygen-free atmosphere glove box because it is air-sensitive.

**Characterization**

Chemical compositions of LaTe$_{2-x}$Sb$_x$ single crystals were determined by Energy Dispersive Spectroscopy (EDS) and Inductively Coupled Plasma (ICP). EDS mapping images were obtained using an ARM-200F (JEOL, Tokyo, Japan). The thickness of the sample was determined by atomic force microscopy (AFM, Bruker). Selected area electron diffraction (SAED) was performed by using a high-resolution transmission electron microscopy (HRTEM, Tecnai F20 super-twin). Electrical resistivity ($\rho$) and specific heat ($C_p$) were measured through the physical property measurement system (PPMS, Quantum Design).

**In-situ high-pressure measurements**

*In-situ* high pressure experiments were carried out in a diamond anvil cell (DAC) with a 300 nm in diameter. We used the four-wire method to measure the electrical transport properties. Cubic boron nitride (cBN) powders were employed as pressure medium and insulating material. The pressure was calibrated by using the ruby fluorescence method at room temperature each time before and after the measurement. In addition, *in-situ* high-pressure powder x-ray diffraction (PXRD) measurements were performed at beam line BL15U of Shanghai Synchrotron Radiation Facility with wavelength $\lambda=0.6199$ Å.

**Theoretical calculations**

Electronic structures of LaTe$_2$ was studied with the density functional theory (DFT) calculations as implemented in the Quantum ESPRESSO (QE) package.[28] Interactions between electrons and nuclei were described by the RRKJ-type ultrasoft pseudopotentials,[29] which were taken from the PSlibrary.[30,31] Generalized gradient approximation (GGA) of Perdew-Burke-Ernzerhof (PBE) formula[32] was adopted for the exchange-correlation functional. A kinetic energy cutoff of the plane-wave basis was set to 80 Ry. A $20\times20\times10$ **k**-point grid was adopted for the Brillouin zone (BZ) sampling. Gaussian smearing method with a width of 0.004 Ry was used for the Fermi surface broadening. In the calculation of the phonon spectrum at ambient pressure, a smearing factor of 0.008 Ry was used. Lattice constants were fixed to the experimentally measured values and only internal atomic positions were optimized with the Broyden-Flercher-Goldfarb-Shanno (BFGS) quasi-Newton algorithm[33] until the force on all atoms were smaller than 0.0002 Ry/Bohr. Orbital-projected FS of LaTe$_2$ under different pressures were obtained by combining the Vienna *ab initio* Simulation (VASP) package[34,35,36] with post-processing VASPKIT package[37] and visualized by using the FermiSurfer package.[38] Electronic susceptibility $\chi$, whose real and imaginary parts are respectively defined as:

$$\chi'(\boldsymbol{q}) = \sum_{nn'\boldsymbol{k}} \frac{f(\varepsilon_{kn}) - f(\varepsilon_{k+qn'})}{\varepsilon_{kn} - \varepsilon_{k+qn'}}$$

$$\chi''(\boldsymbol{q}) = \sum_{nn'\boldsymbol{k}} \delta(\varepsilon_{kn} - \varepsilon_F)\delta(\varepsilon_{k+qn'} - \varepsilon_F),$$

where $f(\varepsilon_{kn})$ is the Fermi-Dirac distribution function and $\varepsilon_{kn}$ is the energy of band $n$ at vector $\boldsymbol{k}$. Dynamical matrix and electron-phonon coupling (EPC) were calculated within the framework of density functional perturbation theory (DFPT)[39,40] as implemented in QE, and the BZ was sampled with a $4\times4\times2$ **q**-point grid and a $60\times60\times30$ **k**-point mesh, respectively.

## Results and discussion

Shinning LaTe$_{2-x}$Sb$_x$ ($x=0$ and 0.4) single crystals were obtained by chemical vapor transport reactions, as shown in Fig. 1b. The atomic ratios of La: Te: Sb are 1: 1.89: 0 and 1: 1.56: 0.37 as ICP measurements, consistent with that of energy dispersive spectrometer (EDS) analysis (Figure S1). Layered LaTe$_2$ can be exfoliated into thin flakes, see Fig. 1c, indicating the inter-layer interactions of LaTe$_2$ is weak. The $\rho$-$T$ curves of LaTe$_{2-x}$Sb$_x$ ($x=0$ and 0.4) from 2 to 300 K are plotted in Fig. 1d. As the temperature decreases, both of $\rho(T)$ curves increase rapidly, indicating two compounds exhibit semiconducting behaviors. The specific heat ($C_p$) of LaTe$_{2-x}$Sb$_x$ ($x$=0 and 0.4) from 2 to 200 K are measured and plotted as a function of temperature in Fig. 1e. The $C_p$ at 200 K are close to the Dulong-Petit value. We fitted the data using the equation $C_p/T=\gamma+\beta T^2$ in the inset of Fig. 1e, and both $\gamma$ are zero, in good agreement with semiconducting behaviors. Meanwhile, the Debye temperatures ($\Theta_D$) are calculated by using the relation $\Theta_D=(12\pi^4 nR/5\beta)^{1/3}$. The values of $\Theta_D$ are 165 K and 179 K. In order to get the temperatures of CDW of LaTe$_{2-x}$Sb$_x$ ($x$=0 and 0.4), we measured the selected area electron-diffraction patterns parallel to the [001] zone axis at room temperature, shown in Figures 1f and g. As the Sb increased, obvious CDW wave vector $q$ changes from $0.5a^*$ to $0.73a^*$, consistent to previous reports.[27] This evidence suggests that both of CDW temperatures of LaTe$_{2-x}$Sb$_x$ are higher than 300K.

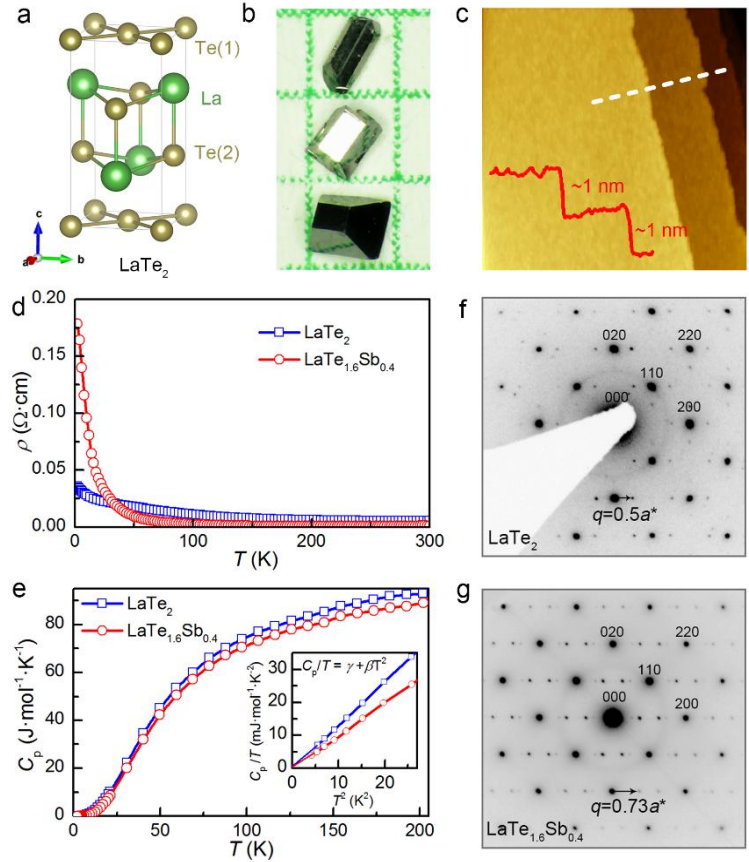

**Figure 1.** (a) Crystal structure of LaTe$_2$. (b) Optical photograph of LaTe$_{2-x}$Sb$_x$ ($x$=0 and 0.4) single crystals. (c) Atomic force microscope topographic image of LaTe$_2$ thin flakes. Step edges with height of 1 nm along the white dot line. The image scale is 200 nm$^2$. (d) Electrical resistivity ($\rho$) as a function of temperature of LaTe$_{2-x}$Sb$_x$ single crystals. (e) Heat capacity ($C_p$) as a function of temperature of LaTe$_{2-x}$Sb$_x$ single crystals. Inset is $T^2$ dependent $C_p/T$. (f, g) Selected area electron diffraction of LaTe$_{2-x}$Sb$_x$ ($x$=0 and 0.4) taken at room temperature along [001] axis. Strong ($hk$0) reflections with $h+k$ even correspond to the sublattice. The CDW wave vectors are $q$=0.5$a^*$ and 0.73$a^*$, respectively.

Figure 2 shows the *in-situ* PXRD patterns of LaTe$_2$ collected at room temperature. At $P$=0.2 GPa, all of the diffraction peaks can be well indexed as the space group of $P$4/nmm. The refinements based on reported structure[19] converge to $R_p$=1.10% and $R_{wp}$=1.58%, see Figure 2a. In Fig. 2b, monotonous shift of all diffraction peaks are observed with the increasing of the external pressure, indicating shrinking of the unit cell. No additional peaks and phase transition emerge below 34.8 GPa. The PXRD pattern of sample after releasing pressure is almost the same as the one of pristine sample. It indicates that $LaTe_2$ does not decompose during measurements. The results of Rietveld refinements for all patterns are summarized in Table S1.

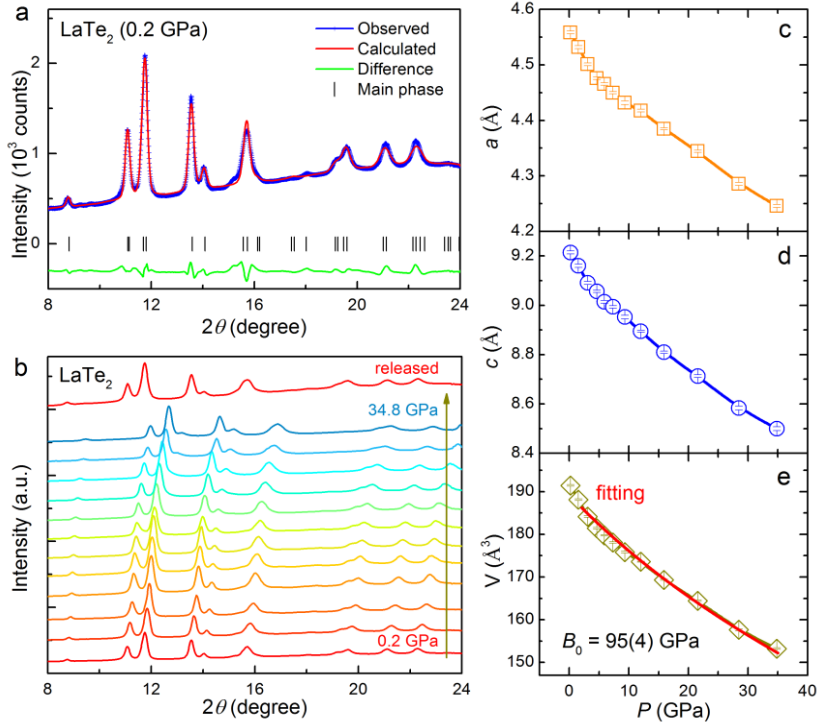

**Figure 2.** (a) Rietveld refinements of *in-situ* PXRD pattern of $LaTe_2$ collected at 0.2 GPa. (b) PXRD patterns of $LaTe_2$ measured under different pressures. (c, d, e) Pressure dependence of the lattice parameters *a*, *c* and *V* of $LaTe_2$. Red line is a fitted curve of Birch-Murnaghan equation.

Next, we have extracted the lattice constants of *a*, *c*, and volume of unit cell (*V*), and put them in Figs. 2c-e. The *a*, *c* and *V* show smooth and continuously decrease with increasing pressure. The whole contraction ratio of $(a_0-a)*100\%/a_0=6.85\%$, $(c_0-c)*100\%/c_0=7.75\%$, $(V_0-V)*100\%/V_0=19.96\%$, where $a_0$, $c_0$ and $V_0$ are the values under ambient pressure. Meanwhile, we fitted the pressure-dependent *V* by using the third-order Birch-Murnaghan equation[41]:

$$P = \frac{3}{2}B[(V_0/V)^{7/3} - (V_0/V)^{5/3}]\left\{1 + \frac{3}{4}(B' - 4)[(V_0/V)^{2/3} - 1]\right\}$$

where *B* is the isothermal bulk modulus, *B'* the first pressure derivative of *B*, *V* and $V_0$ the high-pressure volume and zero-pressure volume, respectively. The fitting curve is plotted in Fig. 2e. The obtained *B* of $LaTe_2$ is 95 (4) GPa, which is higher than $2H_c$-$MoS_2$ (50 GPa)[42] and lower than $2H_a$-$MoS_2$ (110 GPa)[42] and $IrTe_2$ (126 GPa)[43].

Figure 3a shows the temperature dependence of resistance (*R-T* curve) of $LaTe_2$ at various pressures. At low pressures (<4.6 GPa), the *R* increases rapidly upon cooling. It shows typical semiconducting behavior, which is suppressed by increasing pressure. Above 4.6 GPa, the $LaTe_2$ becomes metal, and the values of *R* at 300 K decrease by three orders of magnitude. It demonstrates that $LaTe_2$ undergoes a semiconductor-to-metal transition. Meanwhile, a sudden drop of *R* at 2.8 K is observed at *P*=4.6 GPa, indicating a superconducting transition. As increasing *P* to 20.2 GPa, the $T_c$ increases to the maximal value of 4.6 K. Upon further increasing pressure to 38.0 GPa, the $T_c$ slightly decreases, see Fig. 3b.

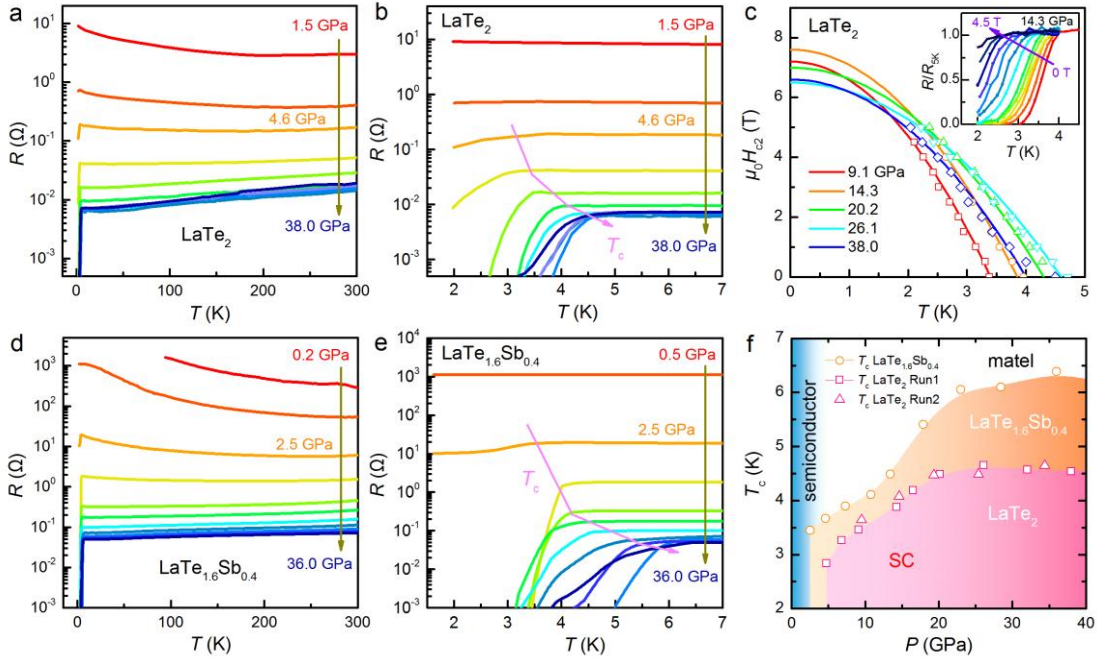

**Figure 3.** (a, b) *R-T* curves of 2-300 K and 2-7 K for $LaTe_2$ under different pressures. (c) Temperature dependence of the upper critical field $\mu_0H_{c2}$ at different pressures of $LaTe_2$. Solid lines are Ginzburg-Landau fitting curves. The inset is $R/R_{5K}$ of 14.3 GPa around $T_c$ under external magnetic field along *c* direction. (d, e) *R-T* curves of 2-300 K and 2-7 K for $LaTe_{1.6}Sb_{0.4}$ under different pressures. (f) Pressure-dependent superconducting phase diagram of $LaTe_{2-x}Sb_x$ (*x*=0 and 0.4).

We measured the magnetic field (*H*)-dependent $T_c$, and put it as inset of Fig. 3c and Fig. S2. All superconducting transitions are gradually suppressed to lower temperature with increasing *H*. The superconductivity completely disappears above 2 K under 6 T. The pressure dependence of $\mu_0H_{c2}(0)$ are plotted in Fig. 3c. Interestingly, the zero-temperature upper critical field $\mu_0H_{c2}(0)$ of 9.1 GPa and 14.3 GPa are estimated to be 7.2 T and 7.6 T, respectively, which are slightly larger than their Pauli-limited critical fields 6.4 T and 7.1 T, respectively.[44] We measured the pressure-dependent $R_H$ and carrier density *n* of $LaTe_2$ at 100 K, shown in Figs. S3-S4. The values of $R_H$ are positive below 1.3 GPa and then turn to negative above 4.1 GPa, revealing the dominant carriers changes from holes to electron. Furthermore, the estimated *n* gradually increases from $10^{20}$ cm$^{-3}$ of 0.8 GPa to $10^{23}$ cm$^{-3}$ of 34.4 GPa, which supports the transition of semiconductor-to-metal.

Figures 3d-e shows the *R-T* curve of $LaTe_{1.6}Sb_{0.4}$ at different pressures. The transitions of semiconductor-to-metal and superconducting of $LaTe_{1.6}Sb_{0.4}$ are also observed. Superconductivity emerges at *P*=2.5 GPa, and the $T_c$ increases to the maximal value of 6.5 K. Accordingly, we summarize all transport results and show a pressure dependence superconducting phase diagram of $LaTe_{2-x}Sb_x$ (*x*=0 and 0.4) in Fig. 3f. With increasing pressure, $LaTe_{2-x}Sb_x$ undergo semiconductor-metal-superconductivity transition at 4.6 and 2.5 GPa, respectively. The highest $T_c$ increases from 4.6 to 6.5 K with the doping of Sb.

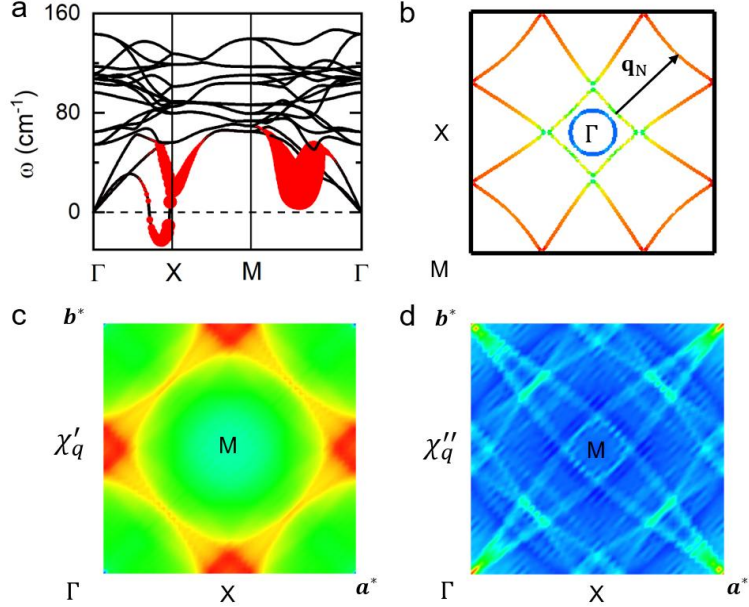

**Figure 4.** (a) Phonon spectrum of LaTe$_2$ at ambient pressure. The size of red dot represents the phonon linewidth $\gamma_{qv}$ of three acoustic branches. (b) Calculated Fermi surface cut in the $k_x$-$k_y$ plane. Color scales on the Fermi surface represent the Fermi velocities, where the red and blue correspond to the highest and zero values, respectively. The real part (c) and imaginary part (d) of the electron susceptibility $\chi$ as a function of ($q_x$, $q_y$) with $q_z$=0 for LaTe2, where the red and blue colors correspond to the highest and lowest values, respectively.

The CDW transition in compound usually occurs when a soft phonon mode at a specific wave vector $q$ emerges. The calculated phonon spectrum of LaTe$_2$ at the ambient pressure is shown in Fig. 4a. There is an imaginary phonon mode near the X point ($q$~0.5$a^*$), in good agreement with the experimentally observed CDW wave vector ($q_{CDW}$~0.5$a^*$). The corresponding Fermi surface cut in the $c^*$=0 plane is displayed in Fig. 4b. There are several quasi-2D Fermi surface sheets across the BZ interior and a small Fermi pocket around the Γ point. Based on the information of the Fermi surface, we calculated the electronic susceptibility $\chi(q)$. The maximum peak in the real part $\chi'(q)$ appears around the X point ($q$~0.5$a^*$), see Fig. 4c, suggesting the instability of the electronic system.[45] On the other hand, the imaginary part $\chi''(q)$ reflects the information of FS nesting.[46] From Fig. 4d, it can be seen that the strongest peak in $\chi''(q)$ locates at the middle point of the Γ-M path, indicating that the nesting vector $q_N$ is around (0.25$a^*$, 0.25$b^*$). In addition, weaker peaks of $\chi''(q)$ show up at the CDW wave vector $q_{CDW}$=0.5$a^*$. There is also a hot spot at the Γ point, which comes from the intraband contribution of a weakly dispersing band, irrelevant to the FS nesting.[47] These facts suggest that the FS nesting alone cannot induce the CDW instability, and the formation of CDW must be assisted by the electron-phonon coupling (EPC) interaction.

To explore the contribution of EPC to the CDW formation, we further calculated the phonon linewidth $\gamma$, which is defined as:

$$\gamma_{qv} = 2\pi\omega_{qv} \sum_{ij} \int \frac{d^3k}{\Omega_{BZ}} \left|g_{qv}(k,i,j)\right|^2 \times \delta(\varepsilon_{q,i} - \varepsilon_F)\delta(\varepsilon_{k+q,j} - \varepsilon_F),$$

where $g_{qv}(k,i,j)$ is the EPC matrix element. The phonon linewidth $\gamma$ takes into account both the effects of the momentum-dependent EPC and the FS nesting. The information of $\gamma$ is overlaid on the phonon spectrum in the form of red dots of Fig. 4a, whose sizes are proportional to their values. It can be seen that there is indeed considerable EPC strength at $q_{CDW}$. Therefore, we conclude that the effect of FS, the strong EPC and the imaginary phonon mode associated with the structural distortions play indispensable roles in inducing the CDW instability.

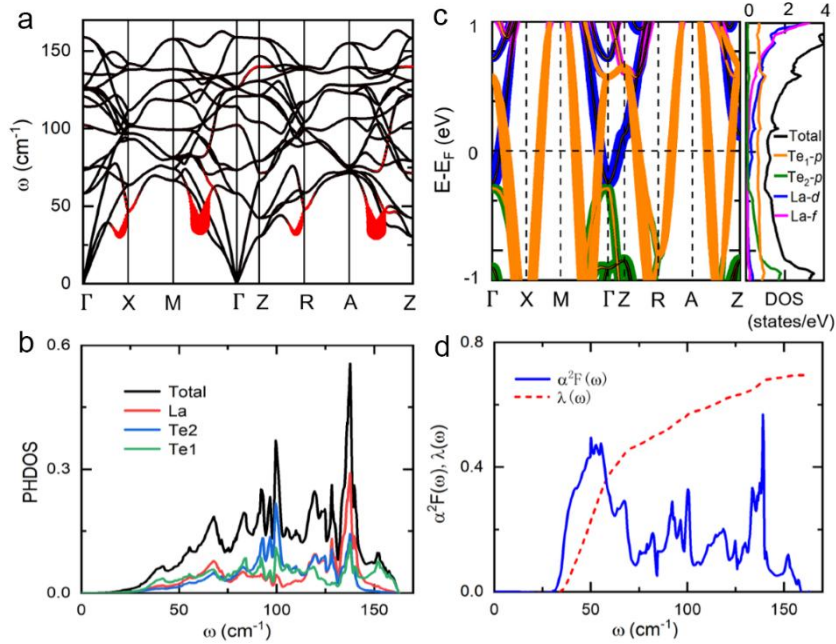

**Figure 5.** (a) Phonon spectrum of LaTe$_2$ under 4.5 GPa. The size of red dot represents the electron-phonon coupling (EPC) strength $\lambda_{qv}$. (b) Total and projected phonon density of states (PHDOS). (c) Orbital-resolved electronic band structure of LaTe$_2$ under 4.5 GPa. The weights of main orbitals are displayed. Right panel shows the total and partial density of states. (d) Eliashberg spectral function $\alpha^2 F(\omega)$ (blue line) and integrated EPC constant $\lambda(\omega)$ (red dashed line).

As we know, the CDW state competes or coexists with other electronic ordered states like superconductivity.[48,49,50] In the CDW-bearing LaTe$_2$, we find experimentally that with the increase of pressure, the superconductivity emerges at $P$=4.5 GPa. In order to explore the superconducting mechanism, we carried out the first-principles calculations on the lattice dynamics, electronic structure, and EPC strength of LaTe$_2$ at 4.5 GPa. The phonon spectrum of LaTe$_2$ at 4.5 GPa is plotted in Fig. 5a. We find that there is no imaginary phonon mode in the whole BZ, indicating that the CDW has been completely suppressed at 4.5 GPa. The corresponding projected phonon DOS (PDOS) is shown in Fig. 5b. It can be seen that the low frequency bands are dominated by the vibrations of Te(1) and La atoms, while the high frequency modes mainly comes from Te(2) and La vibrations. Figure 5c show orbital-resolved electronic band structure of LaTe$_2$ under 4.5 GPa. One can see that two band cross the Fermi level $E_F$ from -1 eV to 1 eV, indicative of metallic behavior. Based on the orbital analysis, these bands are majorly contributed by the 5$p$ orbitals of Te(1) and 4$d$ orbitals of La. And the contribution of Te(1) 5$p$ is larger than that of La 4$d$. From the calculated Fermi surface of LaTe$_2$ under different pressure, shown in Fig. S5, we find that there is a charge transfer from Te(1)-5$p$ to La-5$d$ states and the spherical Fermi pocket dominated by La-5$d$ states at the BZ center enlarges with the applied pressure. From the momentum- and mode-resolved EPC parameter $\lambda_{qv}$ (as indicated by the size of red dots in Fig. 5a), we find that the largest contribution to the EPC comes from the acoustic phonon branch, which corresponds to the sharp peak around 50 cm$^{-1}$ in $\alpha^2 F(\omega)$ (Fig. 5d). The calculated total EPC constant $\lambda$ of LaTe$_2$ at 4.5 GPa is 0.717 and the calculated $T_c$ is about 2.97 K, which is in good accordance with the measured value. This means that the superconductivity in LaTe$_2$ under pressure can be explained in the framework of the BCS theory.

The discovery of pressure-induced superconductivity in LaTe$_2$ have led to a deeper understanding of the origin of superconductivity in the RETe$_2$ system. Firstly, both of LaTe$_2$ and CeTe$_2$ undergo semiconductor-metal-superconductivity transition under pressure. Superconductivity of the former appears at 4.6 GPa with a maximum $T_c$ of 4.5 K. The latter superconductivity occurs at 0.1 GPa and is more sensitive to pressure. It is noteworthy that, with the increasing pressure, the $T_c$ of CeTe$_2$ increases with the increasing of the magnetic phase transition temperature.[22] Secondly, according to the previous calculation of electronic structure, it is suggested that increased self-doped Te(1) 5$p$ hole carriers are responsible for the pressure-induced superconductivity in CeTe$_2$, but the complex magnetic orders make superconductivity mechanism still difficult to understand.[23] The calculation $T_c$ of LaTe$_2$ under pressure in this work is in good accordance with the measured value, which can suggest that the electron-phonon coupling is responsible for pressure-induced SC in RETe$_2$ systems, rather than magnetism. Thirdly, by doping Sb to adjust $q_{CDW}$, the $T_c$ of LaTe$_2$ can enhance to 6.5 K, which is the maximum superconducting transition temperature in the RETe$_2$ system.

## Conclusion

In summary, we have studied the *in-situ* high-pressure electrical measurements up to 40 GPa in LaTe$_{2-x}$Sb$_x$ (*x*=0 and 0.4), together with the high-pressure x-ray diffraction and first-principles calculations. Pressure-induced semiconductor-metal-superconductivity transition happens. With the doping of Sb, the highest $T_c$ increases from 4.6 to 6.5 K. Theoretical calculations demonstrate that the origin of CDW is not only determined by Fermi surfaces nesting, and the appreciable EPC is also important. Meanwhile, EPC is dominated by the low-frequency phonon modes of Te(1) and La atoms, where the former one has a relatively larger contribution. This work establishes significant evidence that the origin of pressure-induced SC is directly related to EPC in RETe$_2$ systems.

## Acknowledgement

This work is financially supported by the National Key Research and Development Program of China (No. 2017YFA0304700, 2018YFE0202601, 2017YFA0302903 and 2018YFA0704300), the National Natural Science Foundation of China (No. 51922105, 11804184, 11974208, 11774424, 12174443, U1932217 and 11974246), and Beijing Natural Science Foundation (Grant No. Z200005). The HP-XRD was performed at the beamline BL15U1, Shanghai Radiation Facility (SSRF). Computational resources have been provided by the Physical Laboratory of High Performance Computing at Renmin University of China.

†These authors contributed equally
Corresponding authors:
kliu@ruc.edu.cn
jgguo@iphy.ac.cn
chenx29@iphy.ac.cn